\documentclass[%
 reprint,
 superscriptaddress,
preprintnumbers,
nofootinbib,
 amsmath,amssymb,
 aps,
]{revtex4-1}

\usepackage{amsmath}
\usepackage{amssymb}
\usepackage{amsthm}
\usepackage{nicefrac}
\usepackage{amsfonts}
\usepackage{slashed}

\usepackage[markup=nocolor]{changes}
\usepackage{natbib}

\usepackage{graphicx}

\usepackage{dcolumn}
\usepackage{bm}

\usepackage{color}

\usepackage[normalem]{ulem}

\newcommand{\bea}{\begin{eqnarray}}
\newcommand{\eea}{\end{eqnarray}}
\newcommand{\be}{\begin{equation}}
\newcommand{\ee}{\end{equation}}

\newcommand{\vecf}{\ensuremath{n}}
\newcommand{\Ggauge}{\ensuremath{g_{\gamma}}}
\newcommand{\Gscalar}{\ensuremath{g_{\phi}}}
\newcommand{\Gfermiono}{\ensuremath{g_{\psi}}}
\newcommand{\Gfermiont}{\ensuremath{h_{\psi}}}

\newcommand{\GN}{\ensuremath{G_{\mathrm{N}}}}

\newcommand{\Lam}{\ensuremath{\Lambda}}

\begin{document}

\title{Indications against dynamical CPT symmetry restoration in quantum gravity}
 
 \author{Astrid Eichhorn}
   \email{eichhorn@thphys.uni-heidelberg.de}
\affiliation{Institute for Theoretical Physics, Heidelberg University, Philosophenweg 16, 69120 Heidelberg, Germany}
\author{Marc Schiffer}
\email{marc.schiffer@ru.nl}
\affiliation{High Energy Physics Department, Institute for Mathematics, Astrophysics, and Particle Physics, Radboud University, Nijmegen, The Netherlands}

\begin{abstract}
CPT symmetry is at the heart of the Standard Model of particle physics and experimentally very well tested, but expected to be broken in some approaches to quantum gravity. It thus becomes pertinent to explore which of the two alternatives is realized: (i) CPT symmetry is emergent, so that it is restored in the low-energy theory, even if it is broken beyond the Planck scale, (ii) CPT symmetry cannot be emergent and must be fundamental, so that any approach to quantum gravity, in which CPT is broken, is ruled out.\\
We explore this by calculating the Renormalization Group flow of CPT violating interactions under the impact of 
quantum fluctuations of the metric. We find that CPT symmetry 
cannot be emergent
and conclude that quantum-gravity approaches must avoid the breaking of CPT symmetry.\\
As a specific example, we discover that in asymptotically safe quantum gravity CPT symmetry remains intact, if it is imposed as a fundamental symmetry, but it is badly broken at low energies if a tiny amount of CPT violation is present in the transplanckian regime. 
\end{abstract}


\maketitle

\emph{CPT symmetry: emergent or fundamental?}\\
It is rare that experimental data can be used to constrain quantum-gravity effects, see \cite{Addazi:2021xuf} for notable exceptions. Here, we leverage strong constraints on CPT violation in the Standard Model (SM) to learn about the properties of quantum gravity at (trans) planckian scales.

Symmetries are at the heart of particle physics. 
One of the most fundamental symmetries of particle physics is CPT symmetry, which combines the three discrete symmetries $C$ (charge conjugation), $P$ (parity transformation) and $T$ (time reversal) into one symmetry operation. 

CPT symmetry is experimentally very well tested \cite{Nodland:1997cc,Anderson:2004qi,Kostelecky:2007zz,Muong-2:2007ofc,Kostelecky:2008be,Kostelecky:2009zp,KLOE:2010yad,MiniBooNE:2011pix,DoubleChooz:2012eiq,Katori:2012pe,KLOE-2:2013ozx,Super-Kamiokande:2014exs,Ding:2016lwt,T2K:2017ega,SNO:2018mge,Escobar:2018hyo,Ding:2019aox,Ding:2020aew,DiDomenico:2020pxd,KATRIN:2022qou,Castulo:2022gpn,Caloni:2022kwp}. This impressive experimental effort has to date not fully been appreciated regarding its implications for quantum gravity. 
Instead, on theoretical grounds it has been argued that quantum gravity effects could  violate CPT symmetry, see, e.g., \cite{Wald:1980nm, Hawking:1985af, Kostelecky:1991ak, Strominger:1993wc, Kostelecky:1995qk, Ellis:1996bv, Mavromatos:2004sz, Mavromatos:2009pp,  Mavromatos:2012ii, Ellis:2013gca, Arzano:2016egk, ValeixoBento:2018tcr, Arzano:2020jro, Addazi:2021xuf, Mavromatos:2022xdo}. This may in fact be expected, if quantum spacetime is discrete, such that concepts of continuum QFT lose their meaning. 

Against this background, an urgent question arises: if quantum gravity breaks CPT, why have the corresponding effects not been detected in experiments?
There are two alternatives: (i) CPT symmetry is broken above the Planck scale, but restored at lower energies, i.e., it is an \emph{emergent} symmetry, (ii) CPT symmetry cannot be emergent and must be imposed at the \emph{fundamental} level. In this case any quantum-gravity theory that breaks CPT symmetry is ruled out.

A symmetry is emergent if all symmetry-violating terms are irrelevant under the Renormalization Group (RG) flow, i.e., they are driven to zero as one evolves the theory from high to low energy scales. A symmetry cannot be emergent, if symmetry-violating terms are RG relevant, because relevant couplings grow under the RG flow to low energies.\footnote{One can still avoid experimental bounds in this scenario by accepting an extreme fine-tuning to tiny values in the transplanckian regime.}
We can thus decide between the two alternatives by calculating loop corrections from quantum gravity to the leading-order CPT-violating terms. \\

\emph{CPT symmetry violation for Standard Model fields}\\
CPT symmetry holds (under some assumptions) in any local, Lorentz invariant quantum field theory. Its violation in perturbatively renormalizable interactions is automatically tied to the violation of Lorentz invariance \cite{Colladay:1996iz,Colladay:1998fq, Greenberg:2002uu}.\footnote{In quantum gravity, at best local Lorentz invariance holds. However, in a regime in which the theory is well-described by small metric fluctuations about a Minkowski background, global Lorentz invariance is in principle recoverable. However, Lorentz symmetry is not dynamically restored in simple settings  \cite{Knorr:2018fdu}.}
Loop corrections to various Lorentz-violating terms have been explored, see, e.g., \cite{Hernandez-Juarez:2018dkx} for matter loops and \cite{Eichhorn:2019ybe} for gravity loops and \cite{Ferrari:2018tps} for a summary.

We will analyze the quantum-gravity effects on the leading-order CPT violating terms for the matter fields in the Standard Model, i.e., for a scalar, fermion and gauge field\footnote{We work in an extension of the Standard Model, where right-handed neutrinos are added, such that all fermion fields of the Standard Model can be combined into Dirac spinors, once we ignore the interactions of the Standard Model, as is sufficient for our analysis.},  which are \cite{Colladay:1996iz,Colladay:1998fq} 
\bea
\label{eq:CPTscal}
S_{\,\mathrm{scalar}}^{\,\mathrm{CPT-odd}}&=&  i\, \Gscalar\int\!\!\!\mathrm{d}^4x\,\sqrt{g} \, \vecf_{\mu}\, \left(\phi^{*}\partial^{\mu}\phi- \phi \partial^{\mu}\phi^{*}\,\right).\\
	\label{eq:CPTgauge}
	S_{\,\mathrm{photon}}^{\,\mathrm{CPT-odd}}&=&\frac{1}{2}\int\!\!\!\mathrm{d}^4x\,\sqrt{g} \, \Ggauge \, \vecf_{\mu}\,\epsilon^{\mu\nu\rho\sigma}A_{\nu}F_{\rho\sigma}\,,\\
	S_{\,\mathrm{fermion}}^{\,\mathrm{CPT-odd}}&=&-\int\!\!\!\mathrm{d}^4x\,\sqrt{g} \,  \vecf_{\mu}\left( \Gfermiono \bar{\psi}\gamma^{\mu}\psi +\Gfermiont \bar{\psi}\gamma^{\mu}\gamma_5\psi\right)\,.\label{eq:CPTfermion}
\eea
Our notation separates a scalar coupling from a vector, such that it relates to the notation in \cite{Colladay:1996iz,Colladay:1998fq} by $g_{\phi}n^{\mu}= \left(k_{\phi}\right)^{\mu}$, $g_{\gamma}n^{\mu}=\left(k_{AF}\right)^{\mu}$, as well as $g_{\psi}n_{\mu}=a_{\mu}$, and $h_{\psi}n_{\mu}=b_{\mu}$.
To explain their form, let us recall that the standard kinetic terms are CPT even and a derivative $\partial_{\mu}$ transforms into $-\partial_{\mu}$ under $PT$ (and is invariant under $C$). Thus, to build a CPT odd term, we substitute one derivative in the kinetic term by an external vector $n_{\mu}$ which is CPT-even. This external vector field violates Lorentz symmetry, because it singles out a preferred frame.
The combinations  $\Gscalar \vecf_{\mu}$, $g_{\gamma}n_{\mu}$, $g_{\psi}n_{\mu}$ and $h_{\psi}n_{\mu}$   parameterize the amount of CPT violation in the theory and are experimentally  tightly constrained.\footnote{Often, constraints are reported on individual components of these interactions; because we are only interested in the order of magnitude of the strongest constraints, we do not make this distinction here. Hence, if we quote bounds on vectors, we refer to the strongest bound on any component of that vector; all details can be found in the corresponding papers.}
For photons, such a term results in birefringence \cite{Kostelecky:2009zp}, which is tightly constrained, e.g., from CMB observations, to $\Ggauge\vecf_{\mu}<10^{-43}\,\rm GeV$ \cite{Kostelecky:2007zz,Kostelecky:2008be,Kostelecky:2009zp}, with more recent constraints \cite{Caloni:2022kwp} stronger by another order of magnitude. Similar constraints ($\Ggauge\vecf_{\mu}<10^{-41}\,\rm GeV$) arise from observation of photons that propagate over astrophysical distances \cite{Nodland:1997cc}. For the Higgs field in the SM, constraints arise from lepton magnetic moments, and constrain $|\Gscalar \vecf_{\mu}|^2 <10^{-29}\, \rm GeV^2$ \cite{Castulo:2022gpn}. In addition, already before the experimental discovery of the Higgs field, constraints were derived based on loop effects of the Higgs field, reading $\Gscalar \vecf_{\mu}< 10^{-29}\, \rm GeV$ \cite{Anderson:2004qi}. For fermions, constraints exist for the various quarks and leptons of the SM. For instance, for the $\tau$, $|n_{\mu}h_{\psi}|< 10^{-10}\, \rm GeV$ \cite{Escobar:2018hyo} and for the $\mu$, $|n_{\mu}h_{\psi}|< 10^{-23}\, \rm GeV$ \cite{Muong-2:2007ofc}.
For stable particles, Penning-trap experiments provide high-precision comparisons of charge-to-mass ratios of particles and antiparticles, resulting in limits on $n_{\mu}h_{\psi}< 10^{-17}\, \rm GeV$ \cite{Ding:2020aew} and even $n_{\mu}h_{\psi}< 10^{-25}\, \rm GeV$\cite{Ding:2016lwt,Ding:2019aox}.
For neutrinos, constraints from tritium-decay amount to $n_{\mu}g_{\psi}<10^{-5}\, \rm GeV$ \cite{KATRIN:2022qou}, and $n_{\mu}g_{\psi}<10^{-20}\, \rm GeV$ from neutrino oscillations \cite{MiniBooNE:2011pix,DoubleChooz:2012eiq,Katori:2012pe,Super-Kamiokande:2014exs,T2K:2017ega,SNO:2018mge}.  In the quark sector, bounds of $10^{-18}\, \rm GeV$ can be derived from Kaon oscillations \cite{KLOE-2:2013ozx,DiDomenico:2020pxd,KLOE:2010yad}.

 Against this impressive experimental effort, the fate of CPT symmetry in quantum gravity becomes a pressing question.\\

\emph{Quantum-gravity fluctuations and CPT restoration}\\
If quantum-gravity effects violate CPT, one would generically expect that the combinations  $g_{\phi}n_{\mu}$ etc.~are all $\mathcal{O}(M_{\rm Planck})\approx 10^{19}\, \rm GeV$, which is clearly many orders of magnitude above the experimental bounds. This is a dimensional estimate which may oversimplify the actual behavior of the system. In particular, if couplings acquire anomalous scaling dimensions due to quantum fluctuations, their values at low energy may be very far from such dimensional estimates. Because the Standard Model is perturbative at energy scales below the Planck scale (with the exception of non-perturbative QCD effects which are irrelevant in the context of the above experimental bounds), quantum fluctuations of matter cannot induce anomalous scaling dimensions which are large enough to solve the tension between experimental bounds and theoretical expectation without finetuning. Indeed, this expectation is bourne out in the explicit study in \cite{Colladay:2007aj}.
We therefore calculate the RG flow of the CPT-violating couplings  under the impact of quantum fluctuations of gravity. These have, to the best of our knowledge, never been calculated before and are the last remaining way to avoid the requirement that the CPT-violating couplings have to be set to values extremely close to zero already in the UV.

There is a seeming ambiguity in this setting, which is the assignment of a scaling dimension to $n_{\mu}$. For dynamical fields, scaling dimensions can be assigned by considering their kinetic terms in the action, which must be dimensionless. However, $n_{\mu}$ in our setting is not a dynamical field, but simply an externally given, fixed field. It does therefore not have a fixed scaling dimension and it is consistent to assign to it either scaling dimension 1, making the couplings $g_{\phi, \gamma,\psi}$ and $h_{\psi}$ dimensionless; or assign to it dimension 0, making the couplings $g_{\phi, \gamma,\psi}$ and $h_{\psi}$ dimensionful, with canonical scaling dimension 1.\footnote{ In settings in which the vector is dynamical, e.g., in Einstein-aether-theory \cite{Jacobson:2007veq}, a dimension can be assigned.}

This ambiguity drops out when we consider the dependence of the combinations $g_i n_{\mu}$ ($i=\phi, \gamma, \psi$, $g_4=h_{\psi}$) under the RG flow, because these always have dimension 1.
For these 
combinations, it holds that
\be
\beta_{g_i n_{\mu}}= k \partial_k\, (g_i n_{\mu}) =  f_i\,  g_i n_{\mu}+ \dots
.\label{eq:betagtilde}
\ee
The factors $f_i$ parameterize the effect of gravitational fluctuations.
We can solve this linear equation (assuming that $f_i \approx \rm const$).
We set initial conditions for the RG flow at a scale $\Lambda_{\rm UV}$, which is a UV cutoff scale beyond which our analysis ceases to be valid.  For instance, for discrete quantum-gravity theories, $\Lambda_{\rm UV}$ would be associated to the discreteness scale. We assume $\Lambda_{\rm UV} \gtrsim M_{\rm Planck}$ and have
\be
(g_{i}n_{\mu})(k)= (g_i n_{\mu})(\Lambda_{\rm UV}) \cdot \left(\frac{k}{\Lambda_{\rm UV}}\right)^{f_i}.\label{eq:scaling}
\ee
The canonical scaling dimension is absorbed into the initial condition and only the anomalous dimension $f_i$ determines whether or not the coupling grows or increases towards the IR.\footnote{We will see below that the situation is slightly more subtle in the case of an asymptotically safe RG fixed point.}

Thus, a discrepancy with the experimental result $g_{i}n_{\mu}\ll  10^{19}
\, \rm GeV$ can only be avoided (i) if the UV value of the coupling is extremely small, $(g_i \, n_{\mu})_{\Lambda_{\rm UV}}\lesssim (g_i \, n_{\mu})_{\rm exp}$ (assuming $f_i \approx 0$), or (ii) if the anomalous scaling dimensions $f_i$ are large enough so that a large value of the coupling at the initial scale is multiplied by a suppression factor much smaller than one. 

Possibility (i) may be realized, if a quantum-gravity theory has a mechanism that requires or at least allows for $g_i\, n_{\mu}(k \simeq M_{\rm Pl}) =0$. We will comment on and explore two such possibilities, namely string theory and asymptotically safe gravity, below. \\
Possibility (ii) may be realized if  $f_i >0$.

\begin{figure}
	\centering
\includegraphics[width=\linewidth,trim={2cm 0 2cm 0},clip]{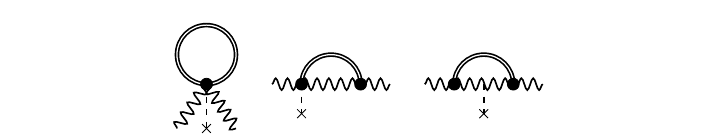}
\caption{Diagrams which contribute to the anomalous scaling dimension of the CPT-violating couplings. The double lines represent metric fluctuations, while the curly line represents any matter field, i.e., scalars, fermions or gauge fields. The dashed line with cross indicates the insertion of the external vector field $n_{\mu}$, which can sit either at the vertex or the propagator. The FRG-regulator can be inserted on either of the internal lines of a given diagram.}
\label{fig:diags}
\end{figure}

Metric fluctuations couple directly to the CPT-odd interaction terms in Eq.~\eqref{eq:CPTscal}-\eqref{eq:CPTfermion} and generate diagrams of the form shown in Fig.~\ref{fig:diags}, which we evaluate with functional RG techniques, see \cite{Wetterich:1992yh, Morris:1993qb,Ellwanger:1993mw, Reuter:1996cp}, and \cite{Dupuis:2020fhh, Saueressig:2023irs, Pawlowski:2023gym} for reviews.  

We approximate the gravitational dynamics by the Euclidean Einstein-Hilbert action (neglecting higher-order curvature terms)
\begin{equation}
\label{eq:EH}
	S_{\mathrm{EH}}=\frac{1}{16\pi \,\GN}\,\int\!\!\mathrm{d}^{4}x\,\sqrt{g}\left(-R+2\Lam\right)\,,
\end{equation}
where $\GN$ and $\Lambda$ are the Newton coupling and cosmological constant, respectively, which we will treat as input parameters. We parameterize metric fluctuations by the dimensionless counterpart of the Newton constant $G = G_N\, k^2$ and the dimensionless cosmological constant $\Lambda = \bar{\Lambda} \,k^{-2}$. Both of these quantities also depend on the RG scale $k$, such that their values that parametrize the metric fluctuations are to be understood as the UV values and not their IR values.

The key result of our FRG computation are the anomalous scaling dimensions for the CPT-violating terms under the impact of gravitational fluctuations,
\bea
f_{\phi}&=& 0+ \mathcal{O}(G^2)\,,\\
f_{\gamma}&=& - G\frac{1- 4\Lambda}{2 \pi\left(1-2 \Lambda \right)^2} \overset{\Lambda \rightarrow 0}{\longrightarrow} - \frac{G}{2\pi}\,,\label{eq:gaugeCPV}\\
f_{\psi}&=&- G \frac{41 - 44\Lambda}{80\pi \left(1-2\Lambda \right)^2} \overset{\Lambda \rightarrow 0}{\longrightarrow} - \frac{41}{80\pi}G\,,\label{eq:fermCPV}
\eea
where the result for the scalar field holds at leading order ($\mathcal{O}(G)$), and we perform a next-to-leading-order study ($\mathcal{O}(G^2)$) in the supplementary material.

\begin{figure}
	\centering
\includegraphics[width=\linewidth]{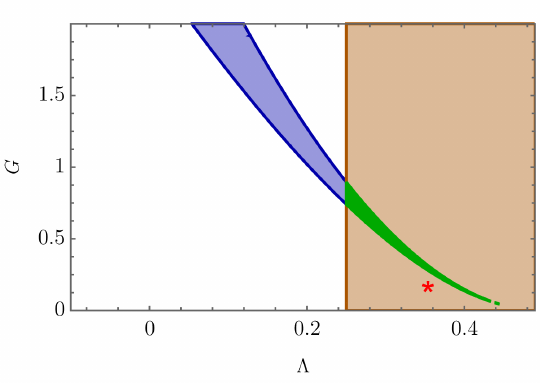}
\caption{Region in the gravitational parameter space, where $f_{\gamma}>0$ (orange), and $f_{\phi}>0$ (blue). We show the small overlap of both regions in green. In this region, CPT symmetry in the scalar and gauge sector is emergent. In contrast, violations of CPT symmetry in the fermion sector still grow towards low energies, even in the green region. The red star indicates the asymptotically safe fixed point for Standard Model matter content \cite{Christiansen:2017bsy}. Its size should not be understood as an estimation of systematic uncertainties; those are expected to be larger.}
\label{fig:gaugescals}
\end{figure}

We thus find the following results, see also Fig.~\ref{fig:gaugescals}:
\begin{itemize}
\item For gauge fields, $f_{\gamma}<0$ in the gravitational parameter space, except for a tiny region. Thus, metric fluctuations generically \emph{increase}, not decrease $n_{\mu}g_{\gamma}$, see supplementary material for studies of the robustness. 
\item For fermions,  $f_{\psi}<0$ for both $g_{\psi}$ and $h_{\psi}$ everywhere in parameter space. Thus, metric fluctuations always \emph{increase}, not decrease the fermionic CPT violating couplings.
\item For scalars, the leading-order contribution to $f_{\phi}$ vanishes. We therefore proceed to next-to-leading order, where we obtain a small contribution that is generically negative. 
\end{itemize}

In summary, we do not find a region in parameter space where metric fluctuations drive the CPT-violating couplings towards zero. Instead, metric fluctuations even increase these couplings further, exacerbating the problem of reconciling experimental constraints with theoretical expectations.
To demonstrate the severity of the required fine-tuning, we consider the observational constraint $g_{\gamma} n_{\mu} < 10^{-43}\, \rm GeV$ \cite{Kostelecky:2007zz,Kostelecky:2008be,Kostelecky:2009zp} and confront it with our result for $f_{\gamma}$, setting $\Lambda=0$ and $G=1$ to obtain a concrete numerical example. For each order of magnitude in scales over which quantum-gravity fluctuations are dynamically important, and where we assume that we can approximate $f_{\gamma} = \rm const$, we obtain an enhancement factor of approximately 1.4. An enhancement is the opposite of the effect that would be needed; in fact, if CPT would be violated in the deep UV, but suppressed by gravitational fluctuations such that observational bounds were met, $f_i \gtrsim 10$ would be necessary to avoid fine-tuning. In contrast, we have $f_{\gamma} \approx - 0.16$ for $G=1$.

We conclude that, under the assumptions underlying our computation, any quantum theory of gravity that violates CPT and Lorentz symmetry can only be reconciled with observations at the cost of extreme fine-tuning of CPT-violating couplings to near-zero at $\Lambda_{\rm UV}$. By adding higher-order CPT-violating operators, this conclusion becomes even stronger: to dynamically restore CPT symmetry at low energies, all of those operators need to be driven to zero by metric fluctuations. As soon as a single operator increases towards lower energies, CPT symmetry cannot be restored without fine-tuning.\\

\emph{The cost of prohibiting CPT violation in asymptotic safety}\\
Asymptotically safe quantum gravity is a quantum field theory of the metric. Based on a proposal by Weinberg \cite{Weinberg:1980gg}, it solves the perturbative nonrenormalizability of General Relativity through an interacting fixed point of the RG. At such a fixed point, all interactions that correspond to perturbative counterterms of General Relativity are generically present. They do not all correspond to independent free parameters, because only those interactions that form relevant perturbations of the fixed point introduce a free parameter. There is not just compelling evidence for the existence of such a fixed point, see \cite{Saueressig:2023irs, Pawlowski:2023gym, Eichhorn:2022gku} for reviews, but also for a total of roughly three free parameters  \cite{Falls:2014tra, Denz:2016qks, Falls:2017lst, Falls:2020qhj}. In addition, long-standing open questions, like the existence of the fixed point in Lorentzian signature as well as the absence of unphysical degrees of freedom, ghosts, are being addressed now, see \cite{DAngelo:2023wje, Saueressig:2025ypi} and  \cite{Knorr:2019atm, Platania:2020knd, Knorr:2021niv,Fehre:2021eob, Platania:2022gtt,Pastor-Gutierrez:2024sbt}, respectively. 
There is also significant evidence for a fixed point when the SM and some of its extensions are coupled to gravity  \cite{Dona:2013qba, Meibohm:2015twa, Biemans:2017zca, Alkofer:2018fxj, Wetterich:2019zdo, Korver:2024sam}, with predictive power for the matter sector \cite{Shaposhnikov:2009pv, Harst:2011zx, Eichhorn:2017ylw, Eichhorn:2017lry, Eichhorn:2018whv}, giving rise to a potential UV completion for the Standard Model \cite{Alkofer:2020vtb, Kowalska:2022ypk, Pastor-Gutierrez:2022nki}. A key property of the asymptotically safe fixed point is its near-perturbative nature, see \cite{Falls:2013bv, Falls:2014tra, Falls:2017lst, Falls:2018ylp, Eichhorn:2018akn, Eichhorn:2018ydy, Eichhorn:2018nda, Kluth:2020bdv}. This implies that calculations that focus on the power-counting relevant and marginal interactions are already producing reliable results in many cases. In the present case, this suggests that there is a leading order of the gravitational contribution to $f$, which consists in gravitational loops; and higher, sub-leading orders, which are based on contributions of power-counting irrelevant couplings.

A second important property is that calculations consistently find the preservation of global symmetries, see, e.g., \cite{Narain:2009fy, Eichhorn:2011pc, Labus:2015ska, Eichhorn:2016vvy, Meibohm:2016mkp, deBrito:2021pyi, Laporte:2021kyp,  Eichhorn:2021qet} (and references therein), contrary to the so-called no-global-symmetries conjecture \cite{Banks:1988yz, Banks:2010zn, Harlow:2018tng, Daus:2020vtf}. This ensures that there is a fixed point at which the CPT violating couplings that we consider vanish, because Eq.~\eqref{eq:betagtilde}. This does not guarantee that they vanish at all scales. If they correspond to RG-relevant parameters, they can depart from the fixed-point value and acquire finite values in the IR. In such a case, it remains a consistent possibility that the couplings are zero, but it is not the only possibility.

In the context of asymptotic safety, the ambiguity regarding the dimensionality of the couplings matters. This is because free parameters are determined by RG relevant couplings. Here it matters whether the vector $n^{\mu}$ has vanishing or non-vanishing mass dimension. For the case of vanishing mass-dimension, the anomalous scaling dimensions $f_i$ must be larger than 1 in order to overwhelm the canonical scaling dimension of the coupling and turn the coupling into an irrelevant coupling that stays at the fixed point for all scales. In contrast, for the case of a non-vanishing mass dimension for $n_{\mu}$, the canonical scaling dimension must only be larger than 0 for the coupling to be irrelevant and thus fixed to zero at all scales.

The most conservative assumption is that $n_{\mu}$ has vanishing mass-dimension and thus the $f$'s must overwhelm a canonical scaling of -1 in order for the CPT-violating couplings to be irrelevant.
Making CPT violation irrelevant in the U(1) gauge sector comes at the cost of a Landau-pole problem in the U(1) sector, at least in the current setup. This is because asymptotically safe gravity can solve the Landau-pole problem, if quantum gravity fluctuations antiscreen the $U(1)$ gauge coupling. There is overwhelming evidence that this happens\footnote{A seeming discrepancy with perturbative calculations has been addressed in \cite{deBrito:2022vbr}.} in most of the gravitational parameter space \cite{Daum:2009dn, Harst:2011zx, Folkerts:2011jz, Christiansen:2017gtg, Eichhorn:2017lry, Eichhorn:2019yzm, DeBrito:2019gdd}. The very small remaining region in which the Landau-pole problem is not resolved is precisely the region in which the CPT-violating coupling becomes irrelevant, see Fig.~\ref{fig:gaugescals}, where we also indicate the location of the asymptotically safe fixed point.\footnote{
The complementarity of the regions is not an accident, but has a structural reason: the beta function of the CPT-violating coupling receives no contributions that renormalize the vertex, because these cancel exactly among each other. The only remaining contribution is to the wave-function renormalization of the gauge field, which, by gauge symmetry, determines the renormalization of the gauge coupling.
}
However, \cite{Christiansen:2017cxa} have shown that a calculation that additionally accounts for the momentum-dependence of the gravitational correction removes this remaining region, thus resolving the Landau-pole problem everywhere and rendering CPT-violation relevant everywhere.

We thus conclude that asymptotically safe gravity likely contains both CPT symmetric theories and CPT-violating theories in its landscape. CPT symmetry remains intact, if the relevant, CPT-violating perturbations of the fixed point are not used in constructing RG trajectories. In other words, it is consistent to \emph{impose} CPT symmetry in asymptotically safe gravity, because it is not generated if CPT violation is set to zero exactly in the UV. However, we do not find a dynamical mechanism that would \emph{explain} why the low-energy theory is CPT symmetric, because this would require CPT-violating interactions to be irrelevant at the fixed point. \\

\emph{Conclusions and outlook}\\
We have discovered that breaking CPT symmetry at the level of dimension-three operators in quantum gravity is only compatible with experimental bounds at the cost of extreme fine-tuning. The already existing problem that a dimension-three operator comes with a coupling of the order of the Planck mass is exacerbated further by quantum-gravity fluctuations \emph{reducing} the scaling dimension of the operator further and thus driving up the scaling dimension of the coupling. Our calculations are subject to approximations, e.g., to leading-order treatment of the gravitational sector. We check the robustness of these approximations in the supplemental material. In addition, we work in Euclidean signature, but expect that our results carry over to Lorentzian signature: within the effective-field-theory setup, gravitational physics is perturbative and the status of an analytical continuation is exactly the same as for non-gravitational QFT; within the asymptotic-safety setup explicit Lorentzian calculations support close agreement with Euclidean ones \cite{Manrique:2011jc,Fehre:2021eob,DAngelo:2023wje,Pastor-Gutierrez:2024sbt,Korver:2024sam,DAngelo:2025yoy,Pawlowski:2025etp,Saueressig:2025ypi}, see also \cite{Biemans:2016rvp,Biemans:2017zca,Knorr:2018fdu, Bonanno:2021squ,Saueressig:2023tfy} for steps towards Lorentzian signature.

Therefore, we conclude that quantum-gravity theories must find a mechanism to preserve CPT symmetry exactly. Examples of such theories exist. For instance, in string theory, CPT symmetry is gauged \cite{Polchinski:1998rr} and must therefore  -- barring anomalies rendering the theory inconsistent \cite{Dine:2004dk} -- stay an exact symmetry, even though it may be broken spontaneously \cite{Kostelecky:1991ak}. More broadly, the no-global-symmetries in quantum gravity has been used to argue for the gauging of CPT \cite{Harlow:2023hjb}. We have discovered a second example here, by finding that in asymptotically safe quantum gravity, a fixed point of the CPT-symmetry-violating terms must lie at vanishing couplings. It is thus consistent to set these couplings exactly to zero at all scales. 

This is not the only possibility consistent with a fixed point, because the CPT-violating terms are RG relevant and can thus depart from the fixed point. There is therefore no predictive power for CPT symmetry in asymptotic safety. The landscape of asymptotically safe theories may include theories with CPT violation. 

Our results imply  -- within the technical limitations of our study -- that any quantum gravity theory which satisfies our assumptions and violates CPT symmetry is likely not phenomenologically viable. This may in particular include theories in which spacetime is discrete and in which therefore P and T are likely not realized. 

Finally, our results are also relevant in the context of proposals for cosmology which do not specify the details of the quantum-gravity era, but are based on CPT symmetry \cite{Boyle:2018tzc}.
\newline\\

\emph{Acknowledgements.}\\
A.~E.~acknowledges helpful discussions with Miguel Montero about symmetries in string theory.
This research was supported by a research grant (29405) from VILLUM Fonden.
 A.~E.~thanks the Perimeter Institute for Theoretical Physics for extended hospitality during the course of this research. The research M.~S.~was in parts supported by the Perimeter Institute for Theoretical Physics. Research at Perimeter Institute is supported in part by the Government of Canada through the Department of Innovation, Science and Economic Development Canada and by the Province of Ontario through the Ministry of Colleges and Universities. The research of M.~S.~was also supported by a Radboud Excellence fellowship from Radboud University in Nijmegen, Netherlands. This
 work is also part of the COST (European Cooperation in Science and Technology) Action CA23130.

\bibliographystyle{apsrev4-2}
\bibliography{references}

\section{Supplemental material}
\section{Details on the setup and methodology}
To extract the scale-dependence of the couplings of the system we employ the functional renormalization group~\cite{Wetterich:1992yh, Morris:1993qb, Ellwanger:1993mw, Reuter:1996cp}, which is based on a scale-dependent effective action $\Gamma_k$, which includes quantum fluctuations of modes with momenta $p^2\gtrsim k^2$. Hence, it interpolates between the classical action, where no quantum fluctuations are integrated out $k\to\infty$, and the full quantum effective action, where all fluctuations are integrated out $k\to0$. The scale-dependence is encoded in the couplings that multiply the interaction terms in $\Gamma_k$. To extract their beta functions, we start from
a differential equation for $\Gamma_k$, which describes how the effective action
changes when lowering the scale $k$. This flow equation reads 
\begin{equation}
\label{eq:floweq}
    k \partial_k \Gamma_k = \frac{1}{2}\text{STr}\left[\left(\Gamma_k^{(2)}+ \mathcal{R}_k\right)^{-1} k \partial_k \mathcal{R}_k\right]\,,
\end{equation} 
where $\mathcal{R}_k$ is the regulator-function, which acts akin a scale-dependent mass term and implements the momentum-shell wise integration of quantum fluctuations. This regulator function has to satisfy a number of conditions, most importantly that $R_k(p^2)>0$ for $p^2<k^2$ and $R_k(p^2)=0$ for $p^2>k^2$. In practise, for bosonic fields we can choose $R_k$ proportional to $\left(k^2/p^2-1\right)\theta(k^2-p^2)$ and for fermions, we can choose it proportional to $\left(\sqrt{k^2/p^2}-1\right)\theta(k^2-p^2)$.
$\Gamma_k^{(2)}$ is a matrix in field space, with the entries given by the second variation of $\Gamma_k$ with respect to all fields. Individual entries of $\Gamma_k^{(2)}$ may also be tensors in spacetime or internal indices. 
Finally,
the $\text{STr}$ indicates a trace over the eigenvalues of the Laplacians that arise in $\Gamma_k^{(2)}$; i.e., about a flat spacetime background, it includes an integration over the loop momentum and a summation over spacetime and internal indices, with an additional sign for fermionic degrees of freedom. For reviews and introductions to the functional renormalization group see \cite{Berges:2000ew, Pawlowski:2005xe, Gies:2006wv, Delamotte:2007pf, Rosten:2010vm, Braun:2011pp, Reuter:2012id, Dupuis:2020fhh, Reichert:2020mja}.

In the current study, we choose the following ansatz for $\Gamma_k$
\begin{equation}
\Gamma_k=\Gamma_k^{\mathrm{grav}}+ \Gamma_k^{\mathrm{matter}}\,,
\end{equation}
where the gravitational part $\Gamma_k^{\mathrm{grav}}$ is
the Einstein-Hilbert action \eqref{eq:EH} supplemented with suitable gauge-fixing:
\begin{equation}
\Gamma_k^{\mathrm{grav}}=S_{\mathrm{EH}}+\frac{1}{ 32 \pi \,G_{\mathrm{N}}\,\alpha_h} \int\!\! \mathrm{d}^4x\,\sqrt{\bar{g}}\, g^{\mu\nu}\,\mathcal{F}_{\mu}\,\mathcal{F}_{\nu}\,,
\end{equation}
with the gauge-fixing condition
\begin{equation}
\mathcal{F}_{\mu}=\left(\delta_{\mu}^{(\alpha}\bar{D}^{\beta)}-\frac{1+\beta_h}{4}\bar{g}^{\alpha\beta}\,\bar{D}_{\mu}\right)h_{\alpha\beta}\,.
\end{equation}
Here, we have performed a decomposition of the full metric into a background $\bar{g}$ and a fluctuation $h$ according to
\begin{equation}
g_{\mu\nu}=\bar{g}_{\mu\nu}+h_{\mu\nu}\,,
\end{equation}
and all barred quantities are with respect to the background metric. In the following, we will choose a flat background, i.e., $\bar{g}=\delta$, and Landau gauge, i.e., $\alpha_h\to0$, but keep $\beta_h$ arbitrary. We highlight that the beta functions we extract do not depend on this choice of background, as one can check explicitly by evaluating the flow equation via background-independent heat-kernel techniques. In practise, working with the specific choice of a flat background is simpler, so we do so here.

For the analysis in the main text we chose $\beta_h=1$, and use residual gauge-dependence of our results as a proxy for the robustness of the qualitative features.\\
For $\Gamma_k^{\mathrm{matter}}$ we choose the standard kinetic terms for scalars, fermions, and gauge fields, supplemented by the CPT-violating terms \eqref{eq:CPTscal}, \eqref{eq:CPTgauge}, and \eqref{eq:CPTfermion}, respectively:
\begin{align}
\Gamma_k^{\mathrm{matter}}=&\frac{1}{2}\int\!\! \mathrm{d}^4x\,\sqrt{g}\left(D_{\mu}\phi^*\,D^{\mu}\phi\right) + S_{\,\mathrm{scalar}}^{\,\mathrm{CPT-odd}}\notag\\
&+\frac{1}{4}\int\!\! \mathrm{d}^4x\,\sqrt{g}\, F_{\mu\nu}\,F^{\mu\nu}+ S_{\,\mathrm{photon}}^{\,\mathrm{CPT-odd}}\notag\\
&\,\,\,\,\,\,+\frac{1}{2\alpha_A}\int\!\! \mathrm{d}^4x\,\sqrt{\bar{g}}\,\left(\bar{D}^{\mu} n_{\mu}\right)\left(\bar{D}^{\nu} A_{\nu}\right)\notag\\
&+\int\!\! \mathrm{d}^4x\,\sqrt{g}\,\bar{\psi}\slashed{D}\psi+ S_{\,\mathrm{fermion}}^{\,\mathrm{CPT-odd}}\,,
\end{align}
where we choose Landau gauge for the Abelian gauge field, i.e., $\alpha_A\to0$.\\
With this ansatz for $\Gamma_k$ we exctract the anomalous dimensions of the matter fields, and the scale-dependence of the CPT-odd couplings by projecting onto the relevant field monomials.

\section{Robustness of $f_{\gamma}<0$}

\begin{figure}[t]
	\centering
\includegraphics[width=\linewidth]{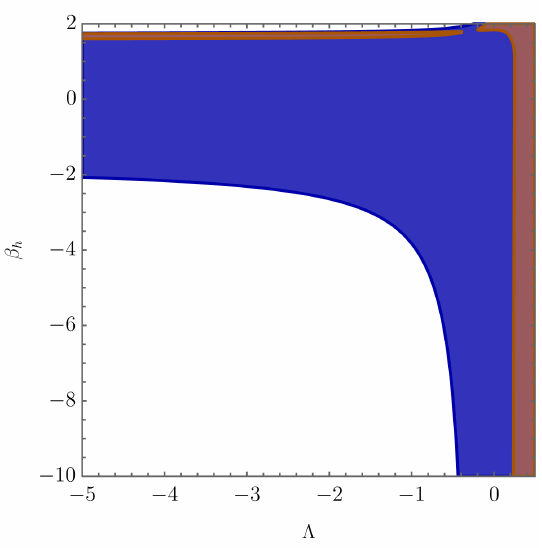}
\caption{Gauge dependence of $f_{\gamma}$. Blue: allowed regions for $\Lambda$, orange: region where $f_{\gamma}>0$. For $\beta_h<1.5$ the orange region is almost gauge-independent.}
\label{fig:fgaugebeta}
\end{figure}

It turns out that $f_{\gamma}$ equals the gravity contribution to the anomalous dimension of the gauge field, which has been computed previously in \cite{Daum:2009dn, Harst:2011zx, Folkerts:2011jz, Christiansen:2017gtg, Christiansen:2017cxa, Eichhorn:2017lry, DeBrito:2019gdd, Eichhorn:2019yzm, Eichhorn:2021qet}, in part including various higher-order effects beyond the ones included here. The result that $f_{\gamma}<0$ is robust, based on these studies; in particular, \cite{Christiansen:2017cxa} shows that by accounting for the momentum-dependence of the gravity-gauge-field- vertex functions, the tiny part of parameter space in which $f_{\gamma}>0$ in our study, also disappears.

We use gauge-dependence as a measure of robustness of the qualitative result that CPT  symmetry in the gauge sector is not emergent. From the diagrammatic structure of CPT-violating contributions, we see that $f_{\gamma}$ is linear in $G$, cf.~Figure~\ref{fig:diags}. Hence, the sign of $f_{\gamma}$ is purely determined by the cosmological constant $\Lambda$ and the gauge-fixing parameter $\beta_h$. At fixed $\beta_h$, only a subset of values for $\Lambda$ are connected to the free fixed point $\Lambda=0$. This is due to poles in the propagator of metric fluctuations, and hence in the flow of the CPT-violating couplings. Therefore, at fixed $\beta_h$ we only allow for those values of $\Lambda$ that are not separated from $\Lambda=0$ by a pole in the propagator. This region is shown in blue in Figure~\ref{fig:fgaugebeta}. In orange we overlay the region where $f_\gamma>0$, i.e., where CPT symmetry is emergent at lower scales. 
This region does not significantly change as a function of $\beta_h$. This holds since the gauge-independent transverse-traceless contributions are dominant over the gauge-dependent scalar contributions of graviton fluctuations. 
These contributions are only dominant for $\beta_h>1.5$, and hence lead to a larger region where $f_{\gamma}>0$.

In summary, $f_{\gamma}<0$ except for a small region at $\Lambda\gtrsim 1/4$ is very robust under changes of the gauge. Furthermore, since the exceptional region at $\Lambda\gtrsim 1/4$ is essentially gauge-independent, we expect that it can be removed completely  at all values of the gauge parameter by taking into account the momentum-dependence of the graviton propagator, as demonstrated in\cite{Christiansen:2017cxa}  for $\beta_h=1$.
\section{Details and robustness of $f_{\phi}<0$}
\begin{figure}[t]
	\centering
\includegraphics[width=\linewidth]{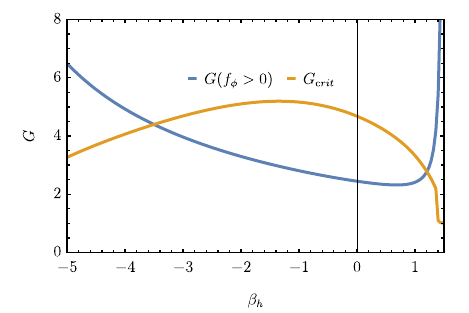}
\caption{Gauge dependence of $f_{\phi}$. The blue line shows the value of $G$ above which $f_{\phi}$ is positive, while the orange line shows where the scalar sector acquires an additional relevant direction, indication the onset of a strong gravity regime.}
\label{fig:fphibeta}
\end{figure}
We find that $f_{\phi}=0$ at leading order for any gauge choice $\beta_h$ and for non-interacting scalars. This is because the diagrams displayed in Figure~\ref{fig:diags} exactly cancel the contribution from the anomalous dimension. Hence, at leading order, there is no gravity-contribution to the CPT-violating scalar interaction. 

To determine whether or not CPT symmetry in the scalar sector can be emergent, we therefore have to include next-to-leading order effect. Here we follow \cite{Eichhorn:2012va, 
deBrito:2023myf} and add scalar self-interactions of the form
\begin{equation}
\Gamma_k^{\mathrm{int}}=\int \mathrm{d}^4x\sqrt{g}\sum_n^{N_{\mathrm{max}}}\left(\frac{K_n}{2k^{4(n-1)}}D_{\mu}\phi^*\,D^{\mu}\phi\right)^n\,,
\end{equation}
which consist of powers of the standard kinetic term, each with a scale-dependent coupling $K_n$. We choose $K_1=Z_{\phi}$ as normalization, and use the beta-functions for $K_n$ to order $N_{\mathrm max}=13$ obtained in \cite{deBrito:2023myf}.\footnote{Specifically, we adapted the results of \cite{deBrito:2023myf}, which were obtained for a single real scalar field by rescaling the pure-matter loops with a factor of two, which accounts for the complex scalar field of our current analysis.} With these interactions, we obtain
\begin{equation}
    f_{\phi}=\frac{K_2}{16\pi^2}\,,\\[-8pt]
\end{equation}
which means that $f_{\phi}>0$ holds for $K_2>0$. General symmetry considerations suggest that interactions like $K_n$ are induced by metric fluctuations, i.e., they cannot be consistently set to zero \cite{Eichhorn:2012va}. 
Since all $K_n$ are (canonically) irrelevant, i.e., IR-attractive couplings, they approach their respective fixed-point value when lowering the energy scale. Therefore, in the current context, we evaluate $K_n$ on their gravity-induced interacting fixed point, which results in $K_{2, *}\sim \mathrm{O}(G^2)$. In \cite{deBrito:2023myf} it was shown that $K_{2, *}$ switches the sign for a
value of $G$ which we call $G(f_{\phi}>0)$ here. For larger values of the Newton coupling, CPT symmetry in the scalar sector can hence be emergent. Additionally, for another value of $G$, $G_{\mathrm{crit}}$, the critical exponent of $K_2$ changes sign, such that the self-interaction itself would become relevant, i.e. IR repulsive and hence a free parameter. We exclude this regime from our analysis, since it requires additional free parameters in the symmetric scalar sector.

These considerations result in the blue region in Figure~\ref{fig:gaugescals}, where $G(f_{\phi}>0)<G<G_{\mathrm{crit}}$, and where $f_{\phi}>0$. Hence, for $\beta_h=1$, there is a small region in the gravitational parameterspace where $f_{\phi}>0$ without introducing new free parameters in the scalar sector.

To study the robustness of this result, we investigate whether this region changes as a function of the gauge-choice. To simplify the analysis, we focus on $\Lambda=0$ and track $G(f_{\phi}>0)$ as well as $G_{\mathrm{crit}}$ as a function of $\beta_h$, which is shown in Figure \ref{fig:gaugescals}.  In the interval $-3.2\lesssim\beta_h\lesssim 1.3$,
$f_{\phi}>0$ can be achieved within a small range of values for $G>1$. 
This interval depends not just on $\beta_h$, but also on $\Lambda$ and may even vanish for some choices.
Overall we conclude that in the scalar sector, CPT symmetry is at best emergent in a small region of the gravitational parameterspace.
\vspace{-12pt}
\section{Robustness of $f_{\psi}<0$}
\begin{figure}[t]
	\centering
\includegraphics[width=\linewidth]{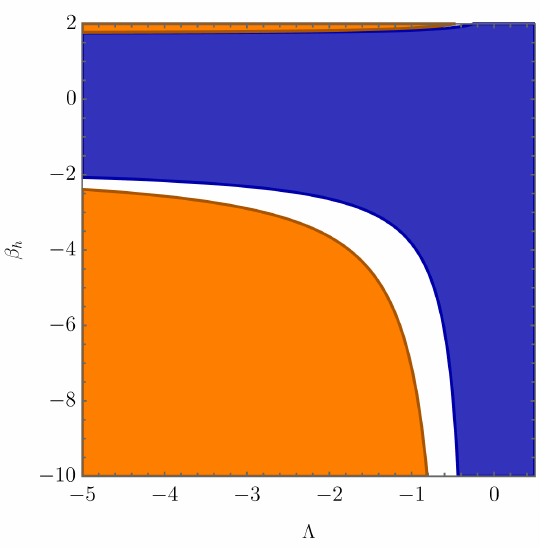}
\caption{Gauge dependence of $f_{\psi}$. Blue: allowed regions for $\Lambda$, orange: region where $f_{\psi}>0$.}
\label{fig:fpsibeta}
\end{figure}

From \eqref{eq:CPTfermion}, which holds for $\beta_h=1$, we see that $f_{\psi}>0$ is only possible for $\Lambda\geq41/44$, which is beyond the pole in the TT-mode of the propagator of metric fluctuations, and hence not relevant for the current study. In Figure~\ref{fig:fpsibeta} we show the region where $f_{\psi}>0$ holds in orange, while we show the allowed range for $\Lambda$ in blue. 
These regions never overlap.
Hence the conclusion that CPT symmetry in the fermion sector is not emergent is gauge-independent, which we interpret as a sign of robustness.

\section{Robustness of under inclusions of higher-order operators}
\begin{figure}[t]
	\centering
\includegraphics[width=\linewidth]{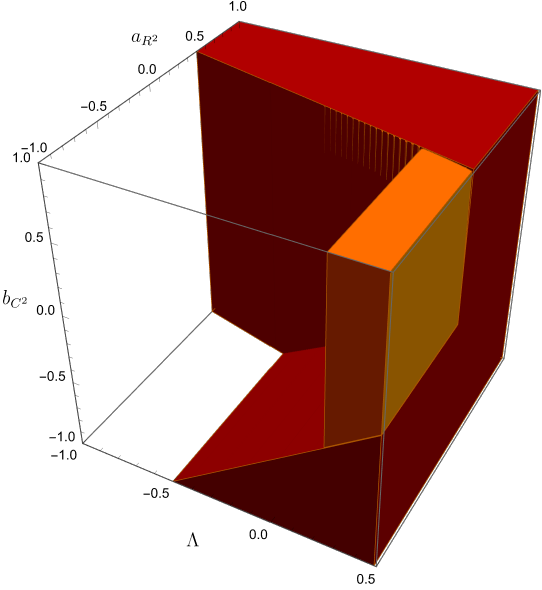}
\caption{Regions in the extended gravitational parameterspace, where $f_{\gamma}>0$ (orange). In red we show the region which lies beyond the propagator poles, and hence needs to be excluded.}
\label{fig:fgauge_HD}
\end{figure}
\begin{figure}[t]
	\centering
\includegraphics[width=\linewidth]{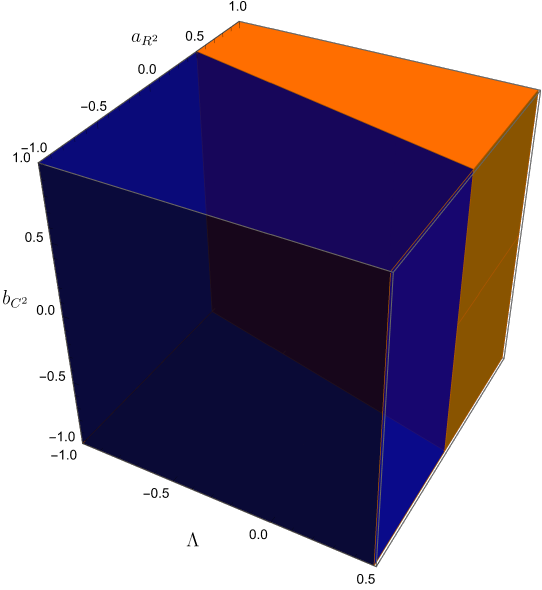}
\caption{ Regions in the extended gravitational parameterspace, where $f_{\psi}>0$ (orange). In blue we show the region which is allowed by the pole-structure of the propagators.}
\label{fig:fferm_HD}
\end{figure}

As a final test for robustness, we add higher-order curvature operators into the system, i.e., we complement $\Gamma_k^{\mathrm{grav}}$ by
\begin{equation}
    \Gamma_k^{\mathrm{HD}}=\frac{1}{16 \pi G_{\mathrm{N}}}\int\!\! \mathrm{d}^4x\,\sqrt{g}k^{-2}\left(a_{R^2} R^2 + b_{C^2} C_{\mu\nu\rho\sigma}C^{\mu\nu\rho\sigma}\right)\,,
\end{equation}
where we have introduce the dimensionless higher-derivative couplings $a_{R^2}$ and $b_{C^2}$. 
The full approximation for the gravitational dynamics is then given by
\begin{equation}
\Gamma_k^{\rm grav} = S_{\rm EH} +   \Gamma_k^{\mathrm{HD}}+ \Gamma_{\rm gauge-fixing}
\end{equation}
We employ standard gauge-fixing, and choose a regulator which is independent of $a_{R^2}$ and $b_{C^2}$. While this makes the momentum integrations more complicated, it ensures that the scale derivative $k\,\partial_k \mathcal{R}_k$ in \eqref{eq:floweq} does not produce occurrences of $\beta_{a_{R^2}}$ or $\beta_{a_{C^2}}$ on the right-hand-side. This allows us to treat $a_{R^2}$ and $b_{C^2}$ as input parameters and investigate the critical exponents of the CPT-violating couplings in the extended parameter-space.

As a first observation, the cancellations observed for Einstein-Hilbert gravity still hold: $f_{\phi}=0$ is true at leading order, even in the presence of higher-derivative couplings. Furthermore, the diagrams directly contributing to $f_{\gamma}$ still cancel, such that  $f_{\gamma}$ is just given by $\eta_A$, as before. 

We now have a four-dimensional parameter space, spanned by $G, \Lambda, a_{R^2}$ and $b_{C^2}$, which we investigate for the gauge choice $\beta_h=1$.  
The beta-functions for $g_{\gamma}$ and $g/h_{\psi}$ are proportional to the Newton coupling. Thus we set $G=1$, and study $f_{\gamma}$ and $f_{\psi}$ in the plane spanned by $\Lambda$, $a_{R^2}$ and $b_{C^2}$.

In Figure \ref{fig:fgauge_HD} we show the region where $f_{\gamma}>0$, i.e., where CPT-violations in the gauge sector are dynamically suppressed. We see that this region is just a natural extension of the region in Einstein-Hilbert gravity. It is bounded at positive $a_{R^2}$ and at negative $b_{C^2}$ by the pole-structure. Furthermore, $f_{\gamma}>0$ is only true for $\Lambda>\frac{1}{4}$, independent of $a_{R^2}$ and $b_{C^2}$. Most importantly, it is still true that this is the region in which the Landau-pole problem is not resolved.

In Figure \ref{fig:fferm_HD} we show the region where $f_{\psi}>0$, i.e., where CPT-violations in the fermion sector are dynamically suppressed. In blue we show the region in the gravitational parameterspace which is viable, as it is connected to $\Lambda=a_{R^2}=b_{C^2}=0$ continuously. We see that these regions do not overlap, as we have observed in the main part for $a_{R^2}=b_{C^2}=0$.

Overall, we see that the addition of higher-order operators does not provide new possibilities for the couplings to be irrelevant; instead, our results from the smaller truncation of the gravitational dynamics are completely robust.

\end{document}